\documentclass[twocolumn]{aastex701}

\usepackage{xspace}
\usepackage{subfigure}
\usepackage{amsmath}
\usepackage{makecell}
\usepackage[super]{nth}

\usepackage{draftwatermark}
\usepackage{xcolor}
\usepackage{xr}
\usepackage{float}

\SetWatermarkText{DRAFT} 
\SetWatermarkScale{7}  
\SetWatermarkColor[gray]{0.95}  

\newcommand{\wow}{Wow!\@ Signal\xspace}

\newcommand*{\HI}{\textsc{Hi}\xspace}

\begin{document}
\raggedbottom

\title{Arecibo Wow! II: Revised Properties of the \wow from Archival Ohio SETI Data}
\author[0000-0002-0786-7307]{Abel M\'endez}
\affiliation{Planetary Habitability Laboratory, University of Puerto Rico at Arecibo}
\email{abel.mendez@upr.edu}

\correspondingauthor{Abel M\'endez}
\email{abel.mendez@upr.edu}

\author[0000-0003-3455-8814,gname=Kevin,sname=Ortiz Ceballos]{Kevin N. Ortiz Ceballos}
\affiliation{Center for Astrophysics ${\rm \mid}$ Harvard {\rm \&} Smithsonian}
\email{kortizceballos@cfa.harvard.edu}

\author[0000-0002-6140-3116]{Jorge I. Zuluaga}
\affiliation{SEAP/FACom, Instituto de F\'{\i}sica - FCEN, University of Antioquia}
\email{jorge.zuluaga@udea.edu.co}

\author[0009-0008-3971-0022]{Kelby D. Palencia-Torres}
\affiliation{Department of Physics, University of Puerto Rico at Rio Piedras}
\email{kelby.palencia@upr.edu}

\author[0000-0002-4772-9670]{Allison J. Smith}
\affiliation{Department of Physics, University of Puerto Rico at Mayagüez}
\email{allison.smith@upr.edu}

\author[0009-0006-2476-0348]{Alondra Cardona Rodr\'iguez}
\affiliation{Department of Astronomy, University of Arizona}
\email{alondrac@arizona.edu}

\author[0000-0001-9896-4622]{H\'ector Socas-Navarro}
\affiliation{European Solar Telescope Foundation}
\email{hector.socas@est-project.eu}

\author[0000-0002-4365-7366]{David Kipping}
\affiliation{Department of Astronomy, Columbia University}
\email{dmk2184@columbia.edu}

\author[0000-0003-4568-2079]{Hodari-Sadiki Hubbard-James}
\affiliation{Department of Physics and Astronomy, Agnes Scott College}
\email{hjames@agnesscott.edu}

\author[0009-0006-2383-6988]{Mai Le}
\affiliation{Department of Physics and Astronomy, Agnes Scott College}
\email{le863@agnesscott.edu}

\author[0009-0000-3417-0794]{Alejandro Rinc\'on-Torres}
\affiliation{School of Engineering, Interamerican University of Puerto Rico at Bayam\'on}
\email{arincon5311@interbayamon.edu}

\begin{abstract}
The \wow, detected in 1977 by the Ohio State University SETI project, remains one of the most intriguing unexplained radio transients. The most recent significant revision of its properties took place in the late 1990s; however, further advances were limited by readily available data from this event. Here we retrieved and analyzed decades of previously unpublished Ohio SETI observations, enabling the most comprehensive re-evaluation of the properties of the \wow to date with modern methods. Our results reveal significant revisions to its parameters that may help explain why its source has been so difficult to identify. We refine its potential origin to two adjacent fields centered on the right ascension $\alpha=19^{\mathrm h}25^{\mathrm m}02^{\mathrm s} \pm 3^{\mathrm s}$ or $19^{\mathrm h}27^{\mathrm m}55^{\mathrm s} \pm 3^{\mathrm s}$, and the declination $\delta=-26^\circ57' \pm 20'$ (J2000), a location both narrower and slightly displaced from earlier estimates. We measure a higher peak flux density exceeding 250~Jy and a frequency of $1420.726 \pm 0.005$~MHz, implying a galactic source with a substantially higher radial velocity than previously assumed. Our analysis provides additional support for the hypothesis that the \wow most likely had an astrophysical origin rather than being attributed to radio interference. In particular, we confirm that small, cold \HI clouds can produce narrowband signals similar to its detection, which might suggest a common origin. These findings provide the most precise constraints to date on the location, intensity, and frequency of the \wow and offer a new path to identify its origin.

\end{abstract}




\section{Introduction \label{sec:intro}} 

The Ohio State University ``Big Ear'' Radio Observatory (OSURO) engaged in a project to search for extraterrestrial intelligence (SETI) from December 1973 until its termination in December 1997. On August 15, 1977, they detected a brief yet intense narrowband radio signal exhibiting attributes expected of a technological transmission originating from outer space. This signal, recorded at a frequency near the hydrogen line (1420 MHz), was famously named the \wow due to astronomer Jerry Ehman's annotation of ``Wow!'' on the data printout \citep{ehman2010big}. Today, it remains an enigmatic and compelling phenomenon in the field of SETI \citep{Tarter2001}. The signal has not been observed again despite extensive subsequent observations \citep{1994Icar..112..485G,gray_vla_2001,gray2002search,harp2020ata,perez2022breakthrough}.

The Big Ear employed an ingenious system consisting of two beam horns, designated as the positive (East) and negative (West) horns. These horns observed the same region of the sky with an approximate interval of three minutes. The \wow was detected exclusively in one of these horns, but the signals from both horns were combined, and it is not known which received the signal. The absence of the signal in both horns suggests that it may have been transient, activated, or deactivated within the three-minute interval between the horns \citep{kipping2022could}. The uncertainty about which of the two horns the signal entered complicates efforts to identify its location and whether it was an artifact of terrestrial interference or an astrophysical phenomenon \citep{gray2002search}.

There are no published research articles in scholarly journals that focused on the detection and analysis of \wow. Its reputation has developed naturally over the decades, primarily through general or conference reports. The first general description of the signal by its discoverers was presented in 1979 in the journal \textit{Cosmic Search} \citep{kraus1979we}. The methods used by Ohio SETI were then described in a conference proceeding \citep{1985IAUS..112..305D}, but it was not focused on \wow. Subsequently, Kraus presented details on the signal in a correspondence with Carl Sagan in 1994 \citep{wowkraus}. Jerry Ehman later authored two comprehensive web articles in 1997 and 2007, and later a book chapter, which provided the current characterization of the properties of \wow \citep{ehman1998big,ehman2010big,ehman2011chapter}.

Most of the research related to \wow has focused mainly on identifying repetitions, while considerably less emphasis has been placed on reevaluating the original signal itself. This tendency can largely be attributed to the insufficiency of available data. Although the data printout illustrating the \wow signal is well known, it is commonly believed that data from the hours or days surrounding the detection were lost since these data are not currently available in the astronomical databases. Regrettably, the Ohio SETI project, as a volunteer initiative, lacked the time and resources to archive its data for future research. The closure of the Big Ear in 1998 was premature, and we all fear that ``anything not saved will be lost'' forever. Fortunately, volunteers from the Big Ear telescope saved most of the data from the Big Ear telescope.

This study offers the first comprehensive reanalysis of the original Ohio SETI data at the time of the \wow detection. Our main objective is to re-evaluate all the properties of the \wow with modern techniques. Our analysis focused on data from 1977 to 1978, although we had to use the data up to the 1990s for validation. Throughout the 1977-1978 interval, the observations generated by the Ohio SETI project remained more consistent.

Our research reveals revised characteristics of the signal, enabling the determination of more precise location, intensity, and frequency. In Section \ref{sec:data}, we provide a detailed description of the Ohio SETI data. Section \ref{sec:transcription} describes the procedures to transcribe and analyze the data. We follow with sections describing refined measurements of the \wow's time (Section \ref{sec:time}), location (Section \ref{sec:location}), flux density (Section \ref{sec:flux}), and frequency (Section \ref{sec:frequency}). Section \ref{sec:alternatives} discusses alternative explanations for the \wow in light of the available evidence.  Finally, our findings are discussed in Section \ref{sec:discussion} and summarized in Section \ref{sec:conclusion}.

\section{The Ohio SETI Data\label{sec:data}}

The Ohio SETI project was active from 1973 to 1997, but not all of its data was recorded in the same format. Data from 1977 to 1984 were preserved by Marc W. Abel from Wright State University. He was among the numerous volunteers instrumental in the operation of Ohio SETI since 1983. This data set is also known as the N50CH record (Number of 50 CHannels) for the name of the software used to process the data. This was also the period when Ohio SETI transitioned from an 8-channel receiver with a chart recorder (the continuum receiver) to a 50-channel 500 kHz receiver using a line printer for data recording (the SETI or spectral receiver). The N50CH software produced the now iconic printout of the \wow.

Between 2006 and 2017, Marc W. Abel carefully photographed over 75,000 printout pages generated by the N50CH software, which he subsequently distributed on a compact disk to various individuals, including Russ Childers from \href{http://www.naapo.org/}{the North American Astrophysical Observatory (NAAPO)}. Russ Childers was another of the volunteers of the telescope in the 1990s who also preserved the last survey of the Ohio SETI data from 1993 to 1997, also known as the Low Budget ETI Search (LOBES) Record. Instead, these data were recorded in digital form.

The N50CH Record was initially stored in PNG and CR2 (Canon Camera) formats with resolutions of 2557 x 1754 and 4416 x 3312, respectively. They were later converted to JPG format at a resolution of 2216 x 1663. The comprehensive data set occupies 1.24 TB of disk space, while the JPG version is significantly reduced to 69 GB. It also includes scans of the IBM 1030 Assembler and FORTRAN IV code used to read and analyze the data, among other ancillary files. Robert Dixon kindly pointed out to us that this data set is available on the NAAPO website managed by Russ Childers. 

For the majority of practical analyses, the resolution offered by the JPG images suffices. They are divided into 167 directories, labeled 001 to 167, plus an additional directory labeled ``wow'' focused on the day of the \wow detection, August 15, 1977. Each directory contains about 300 to 600 images labeled with the directory name and a sequential three-digit number. For example, the first image is labeled 001-001, and the last one is 167-389. The images in the ``wow'' directory are labeled wow-001 to wow-074.

We will use this naming convention, or frame number, to identify the images in the same way as they were originally archived. Most of the data was sequentially archived from 1977 to 1984. An exception pertains to images in directory 001 after frame 001-525, which are from 1978 rather than 1977 as originally indicated, and should be allocated to the 025 directory. Another exception involves the wow directory, which should be positioned between frames 013-411 and 013-412. The last directories also include mixed dates.

\begin{deluxetable*}{ccccccc}
\tablecaption{Summary of Ohio SETI data for August 1977 sorted by the date of the beginning of an observation run, each spanning one or more days. Only three scans were performed at the declination of the \wow, but only the final scan, conducted 48 hours later, covered again its right ascension. Some scans were shorter than one day, as demonstrated on August 22, 1977, or split into two distinct scan, such as August 11 and August 17, 1977. The printout pages can accommodate up to 82 lines, where each line encapsulates approximately 12 seconds of data, or about 16 minutes of data per page.\label{tab:august}}
\tablehead{
    \colhead{Date} & \colhead{Declination} & \colhead{Number of Pages} & \colhead{Start Date} & \colhead{Start Time} & \colhead{End Date} & \colhead{End Time}
}
\startdata
1977-08-02 & -25 40 & 61  & 1977-08-02 & 17:02:53 & 1977-08-03 & 08:41:00 \\
1977-08-05 & -25 40 & 212 & 1977-08-05 & 20:15:57 & 1977-08-08 & 01:05:06 \\
1977-08-08 & -26 00 & 98  & 1977-08-08 & 10:34:08 & 1977-08-09 & 10:44:31 \\
1977-08-09 & -26 00 & 71  & 1977-08-09 & 17:20:40 & 1977-08-10 & 12:18:23 \\
1977-08-11 & -26 20 & 71  & 1977-08-11 & 13:01:09 & 1977-08-11 & 15:48:46 \\
 & & & 1977-08-11 & 18:00:21 & 1977-08-12 & 08:56:29 \\
1977-08-12 & -26 20 & 94  & 1977-08-12 & 08:54:19 & 1977-08-13 & 10:14:56 \\
1977-08-13 & -26 40 & 175 & 1977-08-13 & 10:22:11 & 1977-08-15 & 09:39:33 \\
1977-08-15 & -27 00 & 74  & 1977-08-15 & 13:20:05 & 1977-08-16 & 09:13:24 \\
1977-08-16 & -27 00 & 44  & 1977-08-16 & 09:14:51 & 1977-08-17 & 20:50:45 \\
1977-08-17 & -27 00 & 69  & 1977-08-17 & 07:27:21 & 1977-08-17 & 10:27:23 \\
 & & & 1977-08-17 & 16:51:26 & 1977-08-18 & 08:18:30 \\
1977-08-18 & -27 20 & 89  & 1977-08-18 & 10:52:50 & 1977-08-19 & 09:57:29 \\
1977-08-19 & -27 20 & 73  & 1977-08-19 & 10:12:31 & 1977-08-20 & 03:28:41 \\
1977-08-20 & -27 20 & 77  & 1977-08-20 & 11:05:39 & 1977-08-21 & 08:53:09 \\
1977-08-21 & -27 20 & 69  & 1977-08-21 & 13:38:42 & 1977-08-22 & 07:28:06 \\
1977-08-22 & -27 40 & 4   & 1977-08-22 & 15:11:15 & 1977-08-22 & 15:46:15 \\
1977-08-23 & -27 40 & 76  & 1977-08-23 & 09:31:45 & 1977-08-24 & 08:35:37 \\
1977-08-24 & -27 40 & 76  & 1977-08-24 & 16:59:44 & 1977-08-25 & 13:06:02 \\
1977-08-25 & -27 40 & 75  & 1977-08-25 & 13:29:04 & 1977-08-26 & 09:24:25 \\
1977-08-26 & -28 00 & 104 & 1977-08-26 & 09:37:56 & 1977-08-27 & 12:47:52 \\
1977-08-27 & -28 00 & 166 & 1977-08-27 & 12:52:04 & 1977-08-29 & 09:17:07 \\
1977-08-29 & -28 20 & 83  & 1977-08-29 & 12:42:58 & 1977-08-30 & 10:23:24 \\
1977-08-30 & -28 20 & 27  & 1977-08-30 & 18:57:46 & 1977-08-31 & 00:50:01 \\
1977-08-31 & -28 20 & 60  & 1977-08-31 & 16:37:07 & 1977-09-01 & 09:14:07 \\
\enddata
\end{deluxetable*}

Data collection started on February 5, 1977 with the SMP50 software recording the channel voltages, local sidereal time, and frequency (frame 001-001). On March 11, 1977, the transition to the N50CH software occurred, which recorded the signal-to-noise ratio (SNR) instead of the signal voltage (frame 004-001). The SNR was truncated to positive numbers and represented as numbers from 1 to 9, and then letters from A = 10 to Z = 35. Zero is omitted from the printouts for clarity. Negative standard deviations were introduced on October 22, 1977 (frame 022-198) and are denoted with a strikeout line to the characters. The last available data was recorded on April 25, 1984 (frame 164-311).

Modifications to the analysis procedures and correction of instrumental or software errors were identified and addressed in the N50CH software by the Ohio SETI team. Regrettably, there is an absence of documentation regarding these modifications. Certain changes are evident in the formatting or content of the printouts; however, other alterations became apparent upon numerical verification. Fortunately, an extensive printout of the IBM 1030 Assembler and FORTRAN IV code of the N50CH from circa 1984 is available. This code has facilitated our understanding of some modifications and the interpretation of the procedures used by the Ohio SETI team. Furthermore, we have transcribed and started executing this code using an IBM 1030 computer emulator, replicating the original computational environment. This exercise may yield new information in the future about the \wow or other signals in the data.

The Ohio SETI data are not limited to the N50CH record. Between 1973 and 1977, observations were made using a receiver equipped with four channels, and data was recorded on strip charts. An interruption in data collection occurred from 1984 to approximately 1988, during which the Ohio SETI team upgraded the receiver to a model with additional channels and incorporated the capability of tracking a source for durations of up to an hour.

In the period 1993 to 1997, the LOBES survey was started, during which data were recorded digitally for the first time. LOBES utilized a receiver with 3,000 channels, while the old continuum receiver also recorded data concurrently. Consequently, the Big Ear telescope conducted an all-sky continuum survey using the same instrument over a span of three decades. The spectral SETI survey was conducted for more than two decades and involved two different receivers. These comprehensive and mostly unexplored datasets constitute an invaluable resource for studies in time-domain astronomy.

\section{Data Transcription\label{sec:transcription}}

Our long-term task is the transcription of all Ohio SETI data from physical printouts into a digital format. Our goal is to calibrate and convert these data to align with contemporary astronomical standards. For the purposes of this study, only the data deemed essential for the support of our analysis have been transcribed. So far, we transcribed August 13 to 17, 1977, that is, two days before and after \wow (Table \ref{tab:august}). The printouts display variability in both print and illumination quality; thus, substantial effort is required for data extraction. This is achieved through the use of optical character recognition (OCR), machine learning, and human validation processes.
\setcounter{footnote}{0}
To transcribe the Ohio Big Ear printouts, we opted to use optical character recognition (OCR) tools. We used the OCR software Tesseract\footnote{\href{https://github.com/tesseract-ocr/tesseract}{https://github.com/tesseract-ocr/tesseract}} and trained a model using Tesseract 4.0 neural net training to learn and recognize the printout's numeric characters. The training data set used was one frame wow-053 that had all the numeric characters and it proved to be sufficient for the model to recognize the characters in the printouts. Once the model was trained, a Python pipeline was created using OpenCV\footnote{\href{https://github.com/opencv/opencv}{https://github.com/opencv/opencv}}, an open source Python library for computer vision, to feed our model the characters on the printouts to transcribe. Its output is a text file that contains the data. The transcription process is not perfect, and human verification is still required to confirm and correct any mistakes that the model makes during the process. To minimize the transcription errors made by the model, the printouts must be cleaned before the procedure begins. 

The data from the text files, each for each printout page, were later extracted as comma-separated values (CSV) and joined as a single dataset per day of observation. Subsequently, the data were verified once again with the frames to ensure accuracy. The SNR numerical and alphabetical scale was transformed into cardinal numbers, and the continuum was calculated. The Ohio SETI printouts did not have a continuum column calculated until September 20, 1977 (frame 018-300). Furthermore, we added the object passing the telescope beam from the original Ohio Sky Survey (OSS) catalog \citep{TheMasterList}. This column was originally intended for use, but ultimately remained unused in the Ohio SETI project.

\section{Time\label{sec:time}}

The time column in the printouts shows Eastern Standard Time (EST) for the calendar year, which may not correspond to the local time in Ohio during that period. For example, in August, Eastern Daylight Time (EDT) was applicable in Ohio, so the event detailed in \wow occurred at 10:16 PM EST, equivalent to 11:16 PM EDT in Ohio. The integration period was set to 10 seconds; however, the data was generally recorded every 12 seconds, as an additional 2 seconds were required for the computer to process it. We found many instances where recording occurred after 10 to 15 seconds, though these occurrences were exceptions. 

The solar time recorded in the printouts was derived from the sidereal clock. The determination of the telescope's pointing accuracy in right ascension is critically dependent on the local sidereal time. A variety of astronomical objects might be used to correct for any temporal deviations, such as known radio sources, including the galactic center. Additionally, the Sun and Moon can generate a signal in the telescope as they transit the meridian, even if they are not exactly within the beam field.

During our analysis of the Ohio SETI data, we found evidence that the sidereal clock was offset by seconds after we tried to correct for pointing (Section \ref{sec:location}). It was known that the sidereal clock they were using, the same one they used in the early 1960s, could be off by a few seconds. \cite{ehman2010big} estimated about $\pm$2 sidereal seconds of error at the time of the \wow. To simplify the interpretation of the Ohio SETI data, we chose not to adjust the printed times; instead, we applied the necessary corrections to the calculated coordinates from these times. We only corrected the time for \wow (Table \ref{tab:wow}) based on the fit of the data (Figure \ref{fig:wowfit}).

\section{Location\label{sec:location}}

On the day of the \wow detection, the position of the Big Ear telescope was calibrated using five bright radio sources, as indicated in Table \ref{tab:calibrators}. These sources were chosen because of their brightness, small size, and presence at the \wow declination. In particular, only OC-230.4 (QSO B0118-272) was identified as a calibrator in both the VLA\footnote{\href{https://science.nrao.edu/facilities/vla/observing/callist}{https://science.nrao.edu/facilities/vla/observing/callist}} and Parkes\footnote{\href{https://www.narrabri.atnf.csiro.au/calibrators/}{https://www.narrabri.atnf.csiro.au/calibrators/}} catalog of calibrators. It was verified that the remaining sources did not undergo significant changes in flux over time. This is also corroborated by their flux measurements taken in the early 1970s by the Ohio State Survey (OSS) and subsequently in the 1990s by the NRAO VLA Sky Survey (NVSS).

While the SETI receiver was operating, the 8 MHz continuum receiver was also recording. We obtained the strip-chart data of this receiver for August 16, 1977 from Russ Childers, which was the day after and the same declination of the \wow. This chart clearly displayed the calibration sources because the SETI receiver was not sensitive enough to show them. It included local sidereal time markers at 20-minute intervals (with vertical lines). We calculated their positions using the beam offsets described by \citep{ehman2010big}. These positions were then compared with their current known coordinates.

This analysis revealed a consistent positional offset of approximately 27 seconds on the day of the \wow detection, likely due to a combination of factors, including clock errors. The positive horn squint (offset) described by \citep{ehman2010big} was 154.95 seconds at the declination of the \wow. However, we got an empirically measured value of 128 ± 3 seconds. After applying this correction, we were able to determine the positions of other sources with an accuracy better than three sidereal seconds. Our methodology does not involve fixing the coordinate values documented in the printouts; instead, we recompute them directly on the basis of the printout time.

\begin{figure}
    \centering
    \includegraphics[width=1.0\linewidth]{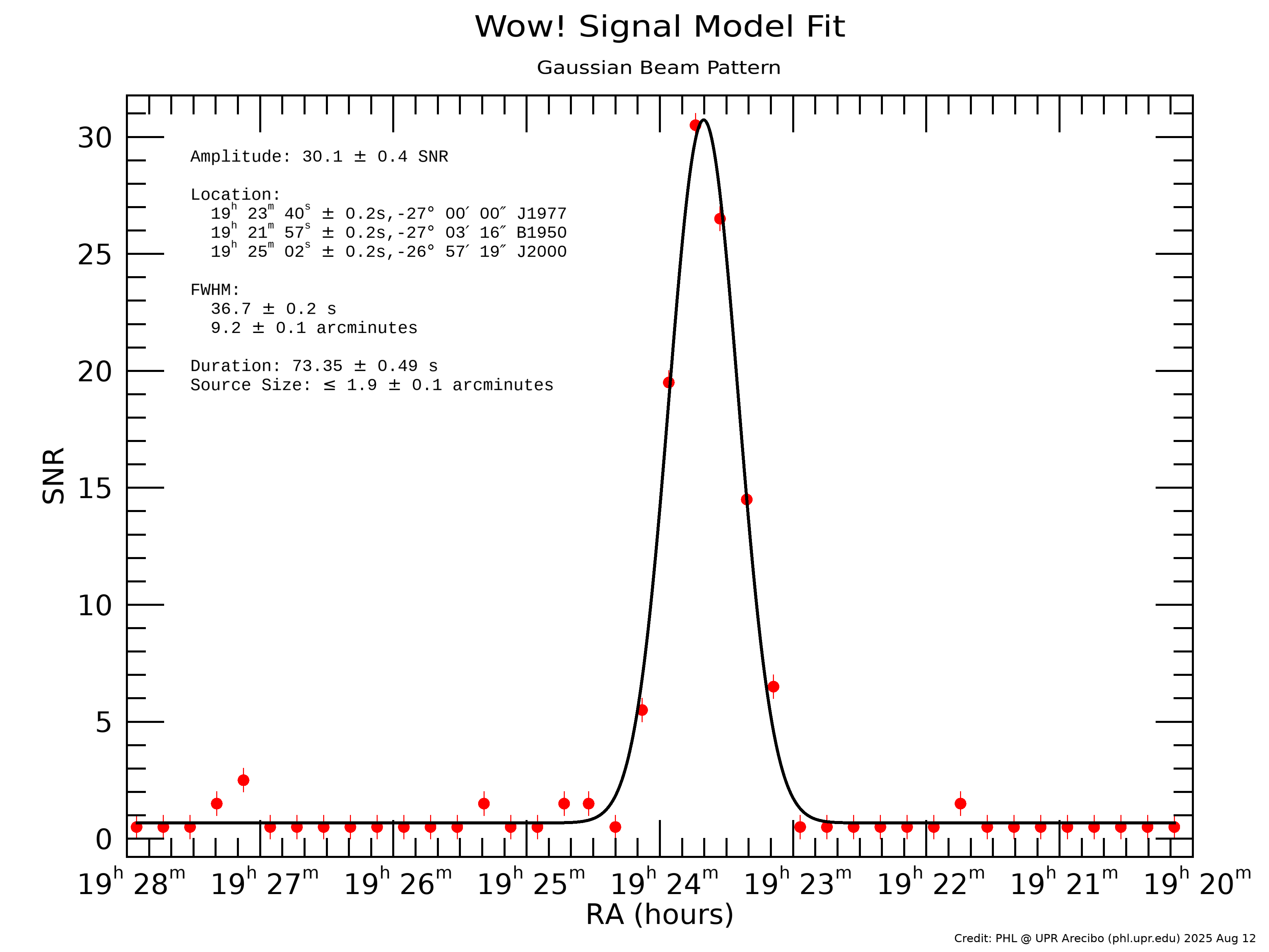}
    \caption{The \wow fitted with a Gaussian beam profile. The peak exhibits a signal-to-noise ratio (SNR) of 30.1$\pm$0.4, centered at the right ascension of 19h 25m 02 $\pm$ 0.2 s (J2000) at the position of the positive horn. The spatial extent of the source is constrained to be less than 1.9 $\pm$ 0.1 arcminutes.}
    \label{fig:wowfit}
\end{figure}

\begin{figure}
    \centering
    \includegraphics[width=1.0\linewidth]{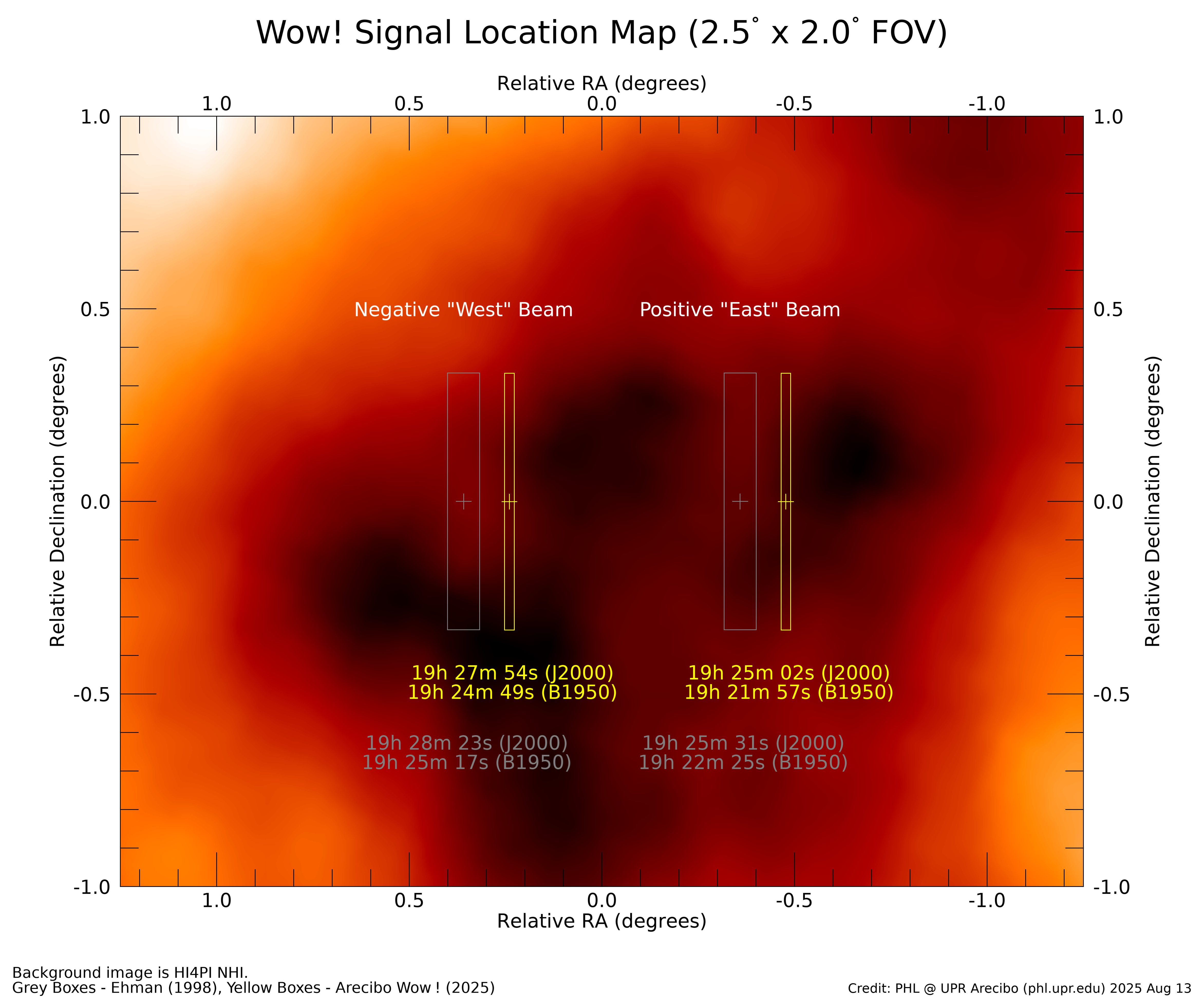}
    \caption{The image shows the possible positions of the \wow in relation to the galactic \HI column density from HI4PI \citep{2016A&A...594A.116H}. Gray boxes indicate the earlier reported positions \citep{ehman1998big}, while the revised positions are marked with yellow boxes, each accompanied by its respective coordinates.}
    \label{fig:wowmap}
\end{figure}

Given that the N50CH time resolution exhibits an uncertainty of $\pm$6 seconds, a Gaussian fit was applied to the \wow to accurately determine its peak position (Figure \ref{fig:wowfit}). The pointing error, quantified at $\pm$3 seconds, significantly exceeds the statistical error of $\pm$0.2 seconds obtained from the fitting process. Together, our calculated error is approximately one third of the previously reported estimates. Consequently, this refinement allows for a more constrained determination of the location of \wow in the right ascension (Figure \ref{fig:wowmap}). The position of the negative horn is calculated from the position of the negative horn as described by \cite{ehman2010big}. In contrast, the same improvement does not apply to declination, as the error persists at $\pm$20 arcminutes. From this analysis, we also obtained the duration and the maximum size of the source, assuming a beamwidth of 8 arcminutes in the right ascension.

\section{Flux Density\label{sec:flux}}

The Big Ear employed two feed horns for the detection of signals, each oriented in slightly different directions. The measured radio signals were converted into an electric potential within the millivolt range. As the Big Ear drift scanned the sky using the Earth's rotation, detected signals would initially appear in the negative horn (also named the West horn) and subsequently, approximately three minutes later, in the positive horn (also known as the East horn). The signal from the negative horn was subtracted from the positive horn, and this difference was recorded at the position of the positive horn \citep{ehman2010big}.

\begin{figure}
    \centering
    \includegraphics[width=1\linewidth]{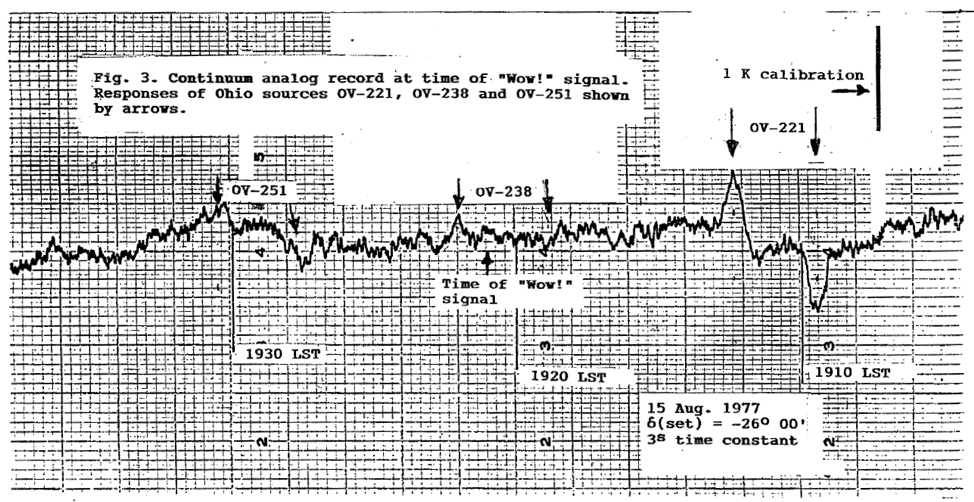}
    \caption{The continuum analog record at the time of \wow, as provided by John Kraus to Carl Sagan in a letter dated 1994. It illustrates three continuum sources, each characterized by negative followed by positive peaks, recorded by the 8-MHz receiver. The temporal axis progresses from right to left, with major grid lines representing two-minute intervals and minor lines representing twenty-second intervals. Note that the \wow signal in a 10 kHz channel would not be perceptible in this plot, as it represents an average across an 8 MHz bandwidth. Figure from \cite{wowkraus}.}
    \label{fig:krausscont}
\end{figure}

Consequently, the positive horn functioned as the ON source, while the negative horn served as the OFF source. Any signal will manifest twice, initially as a negative signal, followed by a positive signal. This pattern is distinctly observable in the continuum record, which was recorded simultaneously with N50CH using an 8 MHz bandwidth receiver (Figure \ref{fig:krausscont}). This is the same receiver used for the OSS in the late 1960s and early 1970s that continued to function until the observatory closed in 1998. Therefore, during all the operations of the Ohio SETI project, this receiver simultaneously and independently recorded the continuum at 1415 MHz, and is a source for calibration and monitoring the performance of the telescope.

All the Ohio SETI data analysis was performed in real time on an IBM 1030 computer and recorded in printouts on an IBM 1132 printer. Therefore, the SNR was determined by an iterative adaptive filtering procedure described in \cite{ehman2010big} as
\begin{equation}
\begin{aligned}
B_i &=
\begin{cases}
\frac{1}{6} S_i + \frac{5}{6} B_{i-1} & \text{if } S_i < S_s \\
B_{i-1} & \text{if } S_i \geq S_s
\end{cases} \\
\sigma_i &=
\begin{cases}
\sqrt{\frac{1}{60} (S_i - B_i)^2 + \frac{59}{60} \sigma_{i-1}^2} & \text{if } S_i < S_s \\
\sigma_{i-1} & \text{if } S_i \geq S_s
\end{cases} \\
\text{SNR}_i &= \frac{S_i - B_i}{\sigma_i}.
\end{aligned}
\label{eq:snr}
\end{equation}
where $S_i$ represents the telescope signal, that is, obtained by subtracting the positive and negative beam signals. The telescope operates with a one-second cadence. However, the SNR is computed at 12-second intervals as the mean of 10 one-second data points and two additional seconds allocated between calculations. The baseline and standard deviation, denoted as parameters $B_i$ and $\sigma_i$, respectively, are derived from prior data points. A predetermined threshold value $S_s$ is used to exclude any prominent signals (\textit{e.g.}, 5 sigma) from the baseline and standard deviation calculations.

Prior to October 22, 1977, the absolute value of $S_i$ was computed in the N50CH record \citep{wowkraus}. Consequently, any constant signal would generate two positive peaks separated by approximately three minutes, with the second peak (the positive horn) associated with its location. Subsequently, in October 1977, the N50CH program was revised to indicate positive and negative signals, similar to the analog record. The \wow was detected before this revision and exhibited only one peak, increasing the ambiguity of its location; specifically, it was uncertain whether this represented the first or second peak and whether the signal was turned on or off during the motion of the horns.

The N50CH sotfware only recorded SNR data, excluding any information pertaining to the flux density calibration of the signals. Jerry Ehman and Russ Childers independently estimated the flux density of the \wow to about 54 or 212 Janskys, respectively \citep{ehman2010big}. To our knowledge, these constitute the only formal estimates of signal strength available, and their derivation is not explicitly documented within the literature. Consequently, to establish a new calibration, we constructed a catalog of OSS sources within the \wow declination ($-27^\circ$ declination) by cross-referencing with the VLA and Parkes catalog of calibrators \citep{1970AJ.....75..351E}. Only OC-230.4 (QSO B0118-272) was cross-referenced in the other catalogs, so we added four additional sources to verify that our flux calibration was consistent along multiple sources (Table \ref{tab:calibrators}).

\begin{deluxetable*}{lllccc}
\tablecaption{List of sources used for the calibration of pointing accuracy and flux density in the Ohio SETI dataset.\label{tab:calibrators}}
\tablehead{
    \colhead{OSS Name} & \colhead{NVSS Name} & \colhead{SIMBAD Identifier} & \colhead{Coordinates (J2000)} & \colhead{OSS Flux (Jy)} & \colhead{NVSS Flux (Jy)}
}
\startdata
OC-230.4 & NVSS 012031-270124 & QSO B0118-272 & 01 20 32 -27 01 25 & 0.84 & 0.934 $\pm$ 0.028\\
OC-266 & NVSS 014127-270607 & PKS J0141-2706 & 01 41 27 -27 06 06 & 1.60 & 1.566 $\pm$ 0.047 \\
OP-282 & NVSS 135210-264927 & QSO B1349-265 & 13 52 10 -26 49 28 & 1.79 &  1.901 $\pm$ 0.057 \\
OF-241 & NVSS 042640-264345 & PKS 0424-26 & 04 26 41 -26 43 55 & 1.05 & 1.220 $\pm$ 0.043 \\
--- & \makecell{NVSS 163139-265644 \\ NVSS 163143-265658} & PKS 1628-268 & 16 31 42 -26 56 51 & --- & 2.230 $\pm$ 0.053 \\
\enddata
\tablecomments{The flux density was measured at 1415 MHz in the Ohio Sky Survey (OSS) and at 1420 MHz in the NRAO VLA Sky Survey (NVSS). PKS 1628-268 comprises two nearby sources, resolved in the NVSS catalog, and their added flux density is shown here.}
\end{deluxetable*}

\begin{figure*}
    \centering
    \includegraphics[width=1\linewidth]{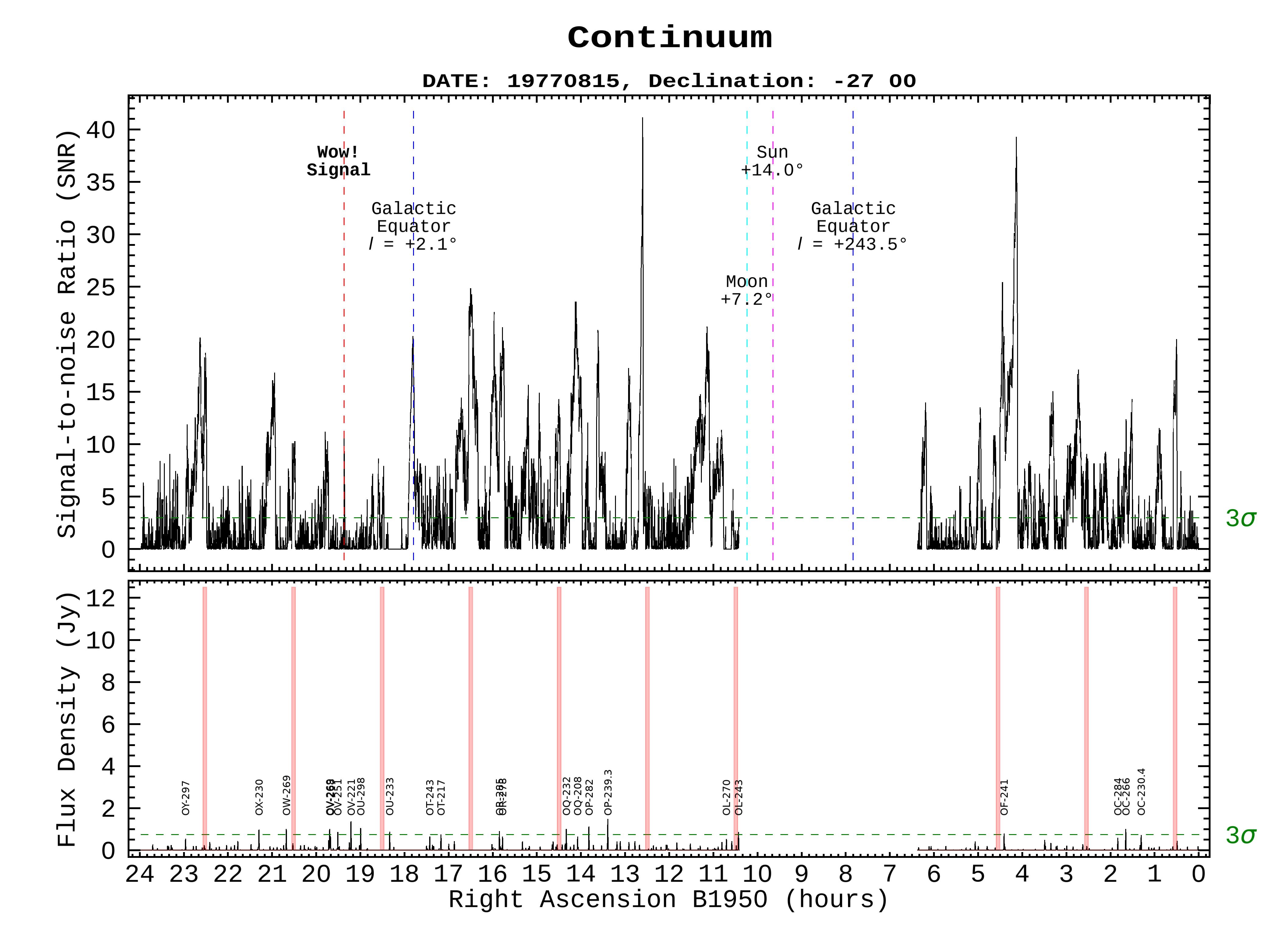}
    \caption{The continuum SNR was derived from the 50-channel output recorded on August 15, 1977 (upper frame). No radio sources located at the specified declination exhibited sufficient strength to be discernible in the SETI receiver (lower frame). Only the signals from the noise tubes (red bars) are evident in the data, displaying variable SNR dependent upon fluctuating noise levels. The intensity of the \wow signal depicted in this plot is diminished due to the continuum being an average of all 50 channels.}
    \label{fig:continuum}
\end{figure*}

A noise tube injected a 5 K calibration signal with a duration of 5 minutes every two solar hours throughout the data acquisition process \citep{cole1976search}. This was the same noise tube utilized during the calibration of the OSS \citep{1968AJ.....73..381D}. Data from the strip chart on August 16, 1977 demonstrated that the noise tube was stable; however, its signal in the SETI receiver exhibited variations attributable to fluctuations in noise levels (Figure \ref{fig:continuum}). A noise tube signal was recorded about one hour before and after the \wow. The initial signal was nearly obscured within the noise levels, as it was close to the noisy galactic center. The subsequent noise tube signal was more distinct, with a peak SNR of 7-sigma. This signal was used to convert the SNR to a flux-density scale using the calibration sources.

First we calculated the noise levels of the continuum in the SETI receiver $\sigma_{\rm cnt}$ (\textit{i.e}, the average of all channels). It is given by
\begin{equation}
\sigma_{\rm cnt} = \frac{S_{\rm ntube}}{\mathrm{SNR}_{\rm ntube}} = \frac{9.4~\text{Jy}}{8.0 \pm 1.8} = 1.2 \pm 0.3~\text{Jy}
\end{equation}
where $S_{\rm ntube}$ is the calibrated flux density of the noise tube from the strip chart and the flux density of OC-230.4, and $\mathrm{SNR}_{\rm ntube}$ is the average SNR of the noise tube signal in the continuum of the SETI receiver. The noise in any of the individual channels $\sigma_{\rm channel}$ is then given by 
\begin{equation}
\sigma_{\rm channel} = \sigma_{\rm cnt} \times \sqrt{n} = 8.5\pm2.1~\text{Jy} 
\end{equation}
where $n$ is the number of receiver channels used in the continuum (50 or fewer, if excluding channels). These noise levels demonstrate why the radio sources located at this declination are not visible in the N50CH dataset, as the majority possess flux densities of less than 1 Jy. Finally, the flux density of the \wow $S_{\rm Wow}$ is
\begin{equation}
S_{\rm Wow} = \mathrm{SNR}_{\rm Wow} \times \sigma_{\rm channel}\geq256\pm63~\text{Jy}
\end{equation}
where $\mathrm{SNR}_{\rm Wow}=30.1\pm0.4$ from the fit analysis shown in Figure \ref{fig:wowfit}.

This result, while higher, remains consistent within the margin of error when compared to the value of 212 Jy determined by Russ Childers in the 1990s \citep{ehman2010big}. The exact noise levels at the moment of the \wow remain uncertain. Had the previous noise tube signal been utilized as a calibrator, the resulting measurement would have been approximately double. Consequently, this represents a lower limit value due not only to the uncertainty in declination but also to uncertainties in the noise levels.

\section{Frequency\label{sec:frequency}}

There are two prior estimations of the frequency of the \wow. The initial estimate, 1420.3556 MHz, was documented in 1994 by John Kraus in correspondence with Carl Sagan \citep{wowkraus}. The subsequent corrected estimate, 1420.4556 MHz, was provided by Jerry Ehman in 1997, after a correction due to an error in which the \nth{1} local oscillator (LO) frequency was 1450.5056 MHz, or 0.1 MHz higher than planned \citep{ehman1998big}. Using the methodology described by \cite{ehman2010big}, we computed the frequency and came to the same result. Subsequently, we expanded the application of this methodology to determine the frequencies corresponding to all channels. Our goal was to search for astronomical sources with known frequencies in the N50CH data to independently validate our methods.

\begin{figure}
\centering
\setlength{\fboxsep}{0pt}    
\setlength{\fboxrule}{0.3pt} 
\fbox{\includegraphics[width=0.45\textwidth]{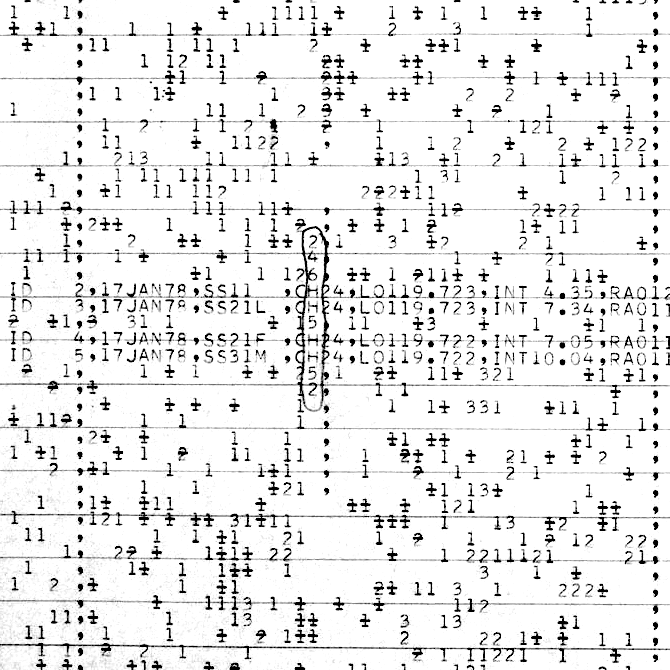}}\\[3mm]
\fbox{\includegraphics[width=0.45\textwidth]{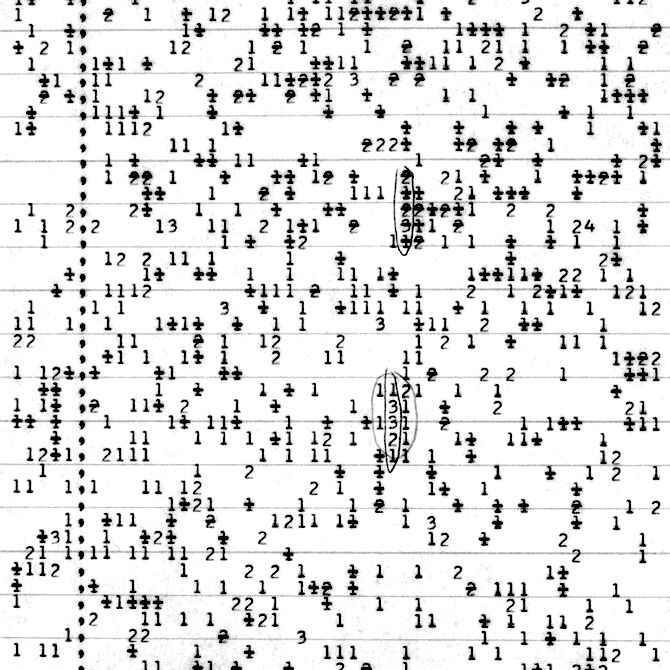}}
\caption{Examples of two narrowband signals in the Ohio SETI data from January 1978, designated Wow2 (top frame) and Wow3 (bottom frame). These signals are similar to the \wow, albeit at a considerably reduced intensity, and in different locations. They were identified and circled here by the Ohio SETI team.}
\label{fig:wows}
\end{figure}

In our analysis, we searched for narrowband signals that were consistently observed across multiple scans, with intervals of days or months. Our analysis concentrated solely on the data collected in 1977 and 1978, bringing us temporally closer to the \wow event. Among the findings, we identified two signals that conformed to the characteristics of the beam pattern, indicative of a small source. We designate these signals, Wow2 and Wow3, due to their similarity to \wow, although they exhibit much weaker intensities (Figure \ref{fig:wows}). It is clear that the Ohio SETI team were interested in these signals, as evidenced by their circling. We believe that they recognized these signals as compact cold \HI clouds, after corroborating their appearance in multiple scans. In fact, the team mentioned such detections in the early days of the Ohio SETI project \citep{1977Icar...30..267D}.

\begin{figure}
\centering
\setlength{\fboxsep}{0pt}    
\setlength{\fboxrule}{0.3pt} 
\fbox{\includegraphics[width=0.45\textwidth]{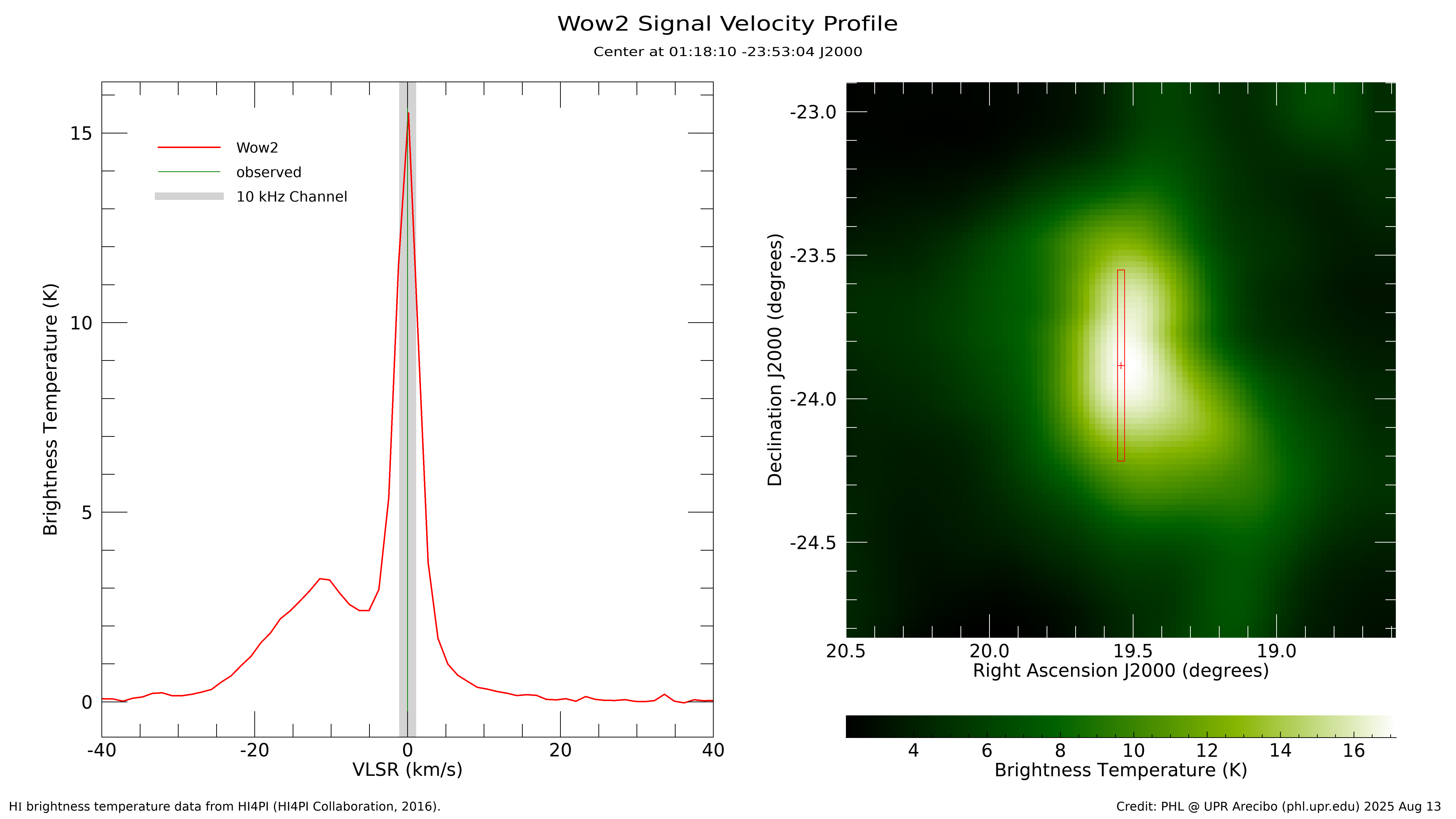}}\\[3mm]
\fbox{\includegraphics[width=0.45\textwidth]{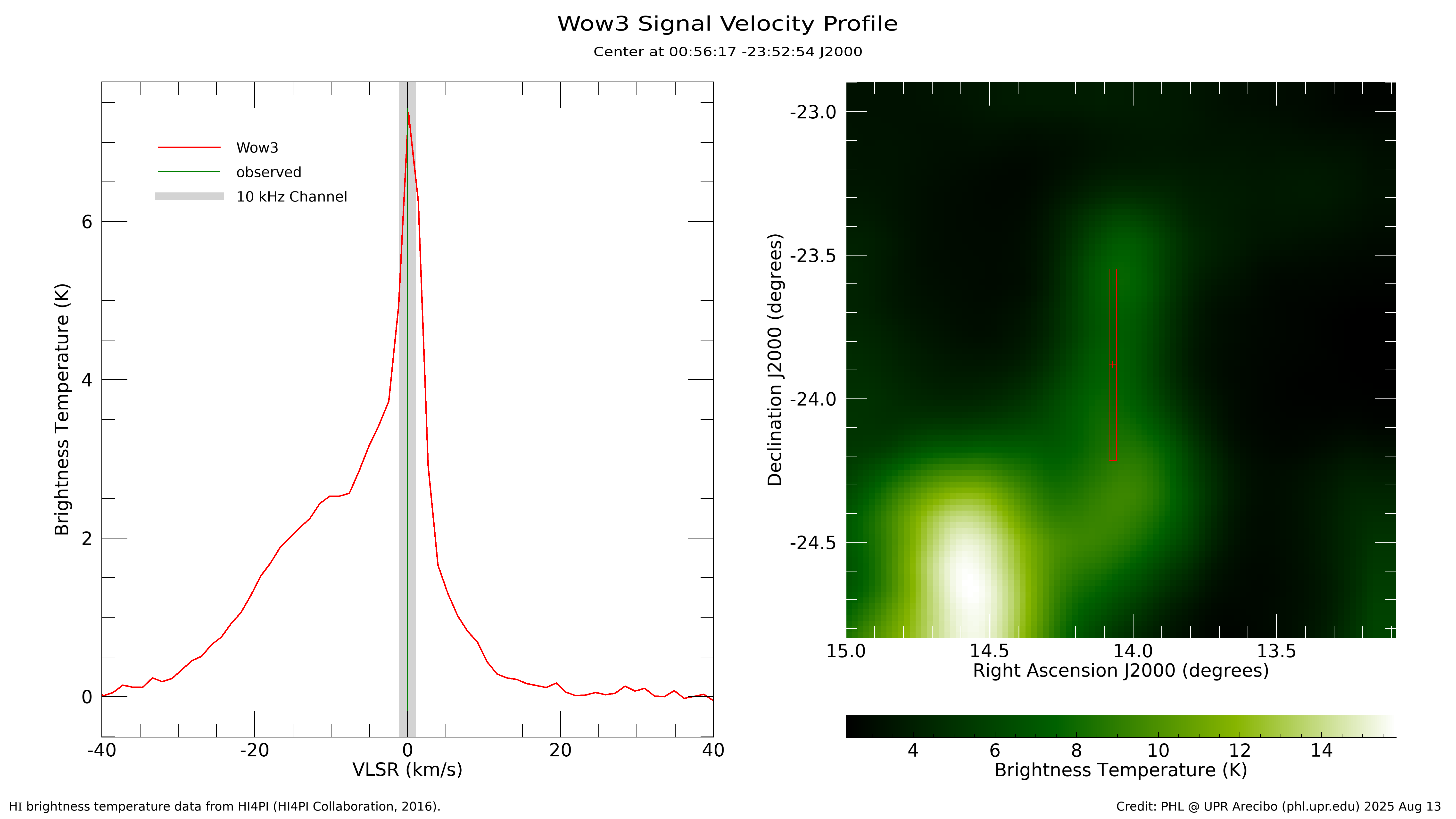}}
\caption{The velocity profiles and brightness temperature maps from HI4PI corresponding to the coordinates of Wow2 (top frame) and Wow3 (bottom frame) provide clear evidence that these two signals are associated with small \HI regions. The velocity profiles observed are also consistent with the expectations for the majority of galactic \HI.}
\label{fig:wowsprofile}
\end{figure}

We searched for the coordinates of these signals in the HI4PI survey and confirmed their association with compact \HI clouds (Figure \ref{fig:wowsprofile}) \citep{2016A&A...594A.116H}. These signals appear to exhibit a narrower frequency range than anticipated from the wider spectral profile of these clouds. This narrowing is not necessarily indicative of lower temperatures, but rather can be attributed to the SETI receivers' insensitivity to the broader lower emissions. We computed the velocity of these clouds relative to the local standard of rest (VLSR) by adjusting their frequency according to Earth's heliocentric motion and the peculiar motion of the Sun. However, we obtained incorrect values for both Wow2 and Wow3, which exceeded the acceptable error margins. Subsequently, we applied the same analytical approach to other broader regions in proximity to the galactic center, and similarly observed incorrect results.

We identified the root cause of the discrepancy as an inversion in channel ordering within the Ohio SETI filter bank, where the frequency decreased with the channel number rather than increased. Although such an ordering can be arbitrary, it diverged from the assumption in the original analysis by \citep{ehman2010big}. By rectifying this inversion, our calculated VLSR aligned perfectly (\textit{i.e.}, VLSR $\approx$ 0 km/s) with that of these clouds maximum \HI emission (Figure \ref{fig:wowsprofile}). Further testing of this methodology across other regions yielded correct results for different latitudes and longitudes of Galactic \HI. It was also discovered that a different correction approach was implemented for the Ohio SETI data acquired before and after December 13, 1977. This implies that there was a software update on that date to adjust the \nth{2} LO frequency from 120.0 MHz to its correct value of 119.9 MHz. Hence, an inversion of frequencies requires a different adjustment of the \nth{2} LO frequency configured before that period.

The computer readout of the N50CH displays the \nth{2} LO frequency instead of the observed frequency, adjusted to the galactic standard of rest (GSR). This indicates that the \nth{2} LO continuously adjusted for the motion of Earth and the Sun relative to a reference frame centered within our galaxy. Any source that is stationary with respect to the galaxy would be observed precisely at the center channel of the SETI receiver (\textit{i.e.}, channel 25.5) with a velocity of zero relative to the GSR. The observed frequency on the central channel $f_c$ and any other channel $f_n$ is determined by
\begin{equation}
\begin{split}
    f_c &= (f_{\text{2LO}} - f_{\text{2LOc}}) + 1420.4056 \ \text{MHz} \\
    f_n &= f_c + (25.5-n) \times 0.010 \ \text{MHz} 
\end{split}
\end{equation}
where $n$ denotes the channel number, ranging from 1 to 50, while $f_{\text{2LO}}$ is the printed \nth{2} LO frequency. The central frequency of the \nth{2} LO is $f_{\text{2LOc}}$, with a value of 120.1 MHz (not 120.0 MHz) prior to December 13, 1977, and 119.9 MHz thereafter. At the time of the central peak of the \wow the \nth{2} LO was 120.185 MHz, which corresponds to a central frequency of 1420.491 MHz and a channel 2 frequency of 1420.726 MHz. The error associated with this measurement is half the bandwidth of the 10 kHz channel.

\section{Alternative Explanations\label{sec:alternatives}}

The Big Ear has been engaged in the search for narrow-band signals for over two decades. Throughout this extensive period, it had ample opportunity to detect astrophysical phenomena, but also random fluctuations, or terrestrial radio frequency interference (RFI), originating either from its internal electronics or external sources.

As part of our search for potential local sources of the \wow, we reviewed news archives and historical documentation from the months immediately preceding and following its detection. This included reports of civil, industrial, or military activities that involved radar or general communications. No events of relevance were identified, and the signal itself was detected on a routine Monday night.

As a historical note, 1977 was also the year two influential science-fiction films, \textit{Star Wars} (May 1977) and \textit{Close Encounters of the Third Kind} (November 1977), were released. Both films reflect the broader an emerging cultural interest in space exploration at the time. This was also the year that the Voyagers spacecraft were launched (August and September 1977). All these events commemorate their 50th anniversary in 2027, together with the \wow.

\subsection{Local RFI}

The \wow represented a notably strong radio signal in comparison to natural background sources at its frequency. Thus, it is plausible to consider that its origin was a terrestrial transmission. The dual horn operation of the telescope would generally cancel out any radio frequency interference occurring simultaneously at both horns, except if it originated from a distant point. The fact that the \wow had the beam pattern implies that the signal originated from a location far beyond Earth’s orbit, appearing stationary within the celestial background. The Moon has the capacity to reflect terrestrial emissions back towards Earth. However, the signal was detected during a New Moon phase at a time when both the Sun and the Moon were on the opposite side of Earth.

Let us consider the hypothesis that the signal was generated by RFI that had the same likelihood of producing both weak and strong signals with durations ranging from fractions of a second to the full 72-second detection window. In this scenario, the three-sigma probability of such an event, ranging from 4 to 35 sigmas (from a significant detection to the scale threshold ``Z''), is calculated as (1/32 possibilities)$^6$ = 1/1,073,741,824, or approximately one in a billion. If this RFI is persistent, it would take approximately six millennia of uninterrupted observations to achieve a signal analogous to the \wow from random RFI occurrences.

Therefore, any terrestrial source as an explanation of the \wow is against very low probabilities unless this source is known a priori to produce signals with a characteristic Gaussian pattern. In contrast, point radio sources are 100\% likely to produce this pattern.

\subsection{Second Harmonic}

One reason the \wow is considered unlikely to be terrestrial RFI is that it was detected at a frequency near 1420 MHz, which is a protected band reserved for astronomical observations. Although this protection reduces the likelihood of intentional transmissions, it does not completely rule out the possibility of illegal or unintentional interference. Another potential explanation involves harmonic generation. Specifically, the \wow observed at 1420 MHz could represent the second harmonic of a strong signal centered around 710 MHz.

The second harmonic of a signal is a frequency that is exactly twice the fundamental frequency. Harmonics are commonly produced in electronic systems due to non-linearities in components such as amplifiers, mixers, and transmitters. When a strong signal passes through such a non-linear device, it can generate spurious emissions at integer multiples of the original frequency, including the second harmonic.

In this context, a powerful signal at 710 MHz could, in principle, produce a harmonic at 1420 MHz if it encountered the right non-linear conditions. This frequency lies exactly at the boundary between UHF TV channels 53 (704 - 710 MHz) and 54 (710 - 716 MHz). We reviewed all UHF television stations operating in Ohio at the time of the signal (August 1977) and found no stations transmitting on either of those channels \citep[Table \ref{tab:ohio_uhf_1977};][]{BroadcastingYearbook1977,TelevisionFactbook1977}.
To be thorough, we also examined a third-harmonic scenario, corresponding to fundamental frequency of approximately 473 MHz, or UHF channel 14 (470 - 476 MHz). Again, no known transmitters were operating at those frequencies in Ohio during that period. Harmonics also tend to be weaker than their fundamental signals, while the \wow was notably strong.

\begin{deluxetable*}{llll}
\tablecaption{Ohio UHF TV stations active around August 1977.\tablenotemark{a} Only channels 14, 53, and 54 were capable of generating harmonics at the hydrogen line frequency, but none were in operation.\label{tab:ohio_uhf_1977}}
\tablewidth{0pt} 
\tablehead{
\colhead{UHF Channel} & \colhead{Call Sign (in 1977)} & \colhead{City of License} & \colhead{Status/Affiliation (1977)}
}
\startdata
17 & WJAN & Canton & Independent/Religious \\
19 & WXIX-TV & Newport, KY & Independent \\
20 & WOUB-TV & Athens & PBS/Educational \\
23 & WAKR-TV & Akron & ABC \\
25 & WVIZ & Cleveland & PBS/Educational \\
34 & WOSU-TV & Columbus & PBS/Educational \\
43 & WUAB & Lorain & Independent \\
45 & WNEO & Alliance & PBS/Educational \\
48 & WCET & Cincinnati & PBS/Educational \\
49 & WEAO & Akron & PBS/Educational \\
57 & WBGU-TV & Bowling Green & PBS/Educational \\
\enddata
\tablenotetext{a}{Data compiled from the \cite{BroadcastingYearbook1977} and the \cite{TelevisionFactbook1977}.}
\end{deluxetable*}

On a curious note, it seemed plausible that the UHF channel layout around 710 MHz might have been designed to minimize the risk of second harmonics interfering with the hydrogen line at 1420 MHz. However, we found that this alignment is purely coincidental, as the UHF channel structure was determined by linear allocation rather than harmonic considerations.

\subsection{Satellites and Space Probes}

Another proposed explanation for the \wow is that it might have originated from a satellite, possibly one in geostationary orbit. While geostationary satellites typically remain fixed over the equator, those that are no longer actively station-kept can drift into inclined geosynchronous orbits, causing them to trace a figure-eight pattern in the sky over the course of a day. However, even in such inclined orbits, the satellite’s apparent motion in declination is limited to the value of its orbital inclination.

In 1977, most geostationary satellites had inclinations well under $15^\circ$, and none exceeded $20^\circ$. To appear at the \wow’s declination of approximately $-27^\circ$, a geostationary satellite would need an inclination of at least $27^\circ$, which is far beyond the inclination of any satellite operating or drifting at the time. Furthermore, even if such a satellite existed, it would not have been visible from the latitude of the Big Ear telescope in Ohio, as the geometry of geosynchronous orbits restricts their visibility to within a narrow band around the celestial equator from mid-northern latitudes.

Another class of satellites worth considering are those in Molniya-type orbits, highly elliptical inclined orbits used extensively by the Soviet Union. These satellites do reach much higher inclinations (about $63^\circ$) and can cross declinations near $-27^\circ$ as they swing through their perigee over the Southern Hemisphere. However, their high speed near perigee causes them to move rapidly across the sky.

A Molniya satellite would spend only a fraction of a second crossing the narrow beam of the Big Ear telescope, whose beamwidth at 1420 MHz corresponds to just a few arcminutes. In contrast, the \wow lasted for the full 72-second window of the telescope’s fixed-beam scan, with a profile consistent with a sidereal drift, the same motion exhibited by distant celestial sources due to Earth’s rotation. The transit time and stability of the signal are inconsistent with the brief, fast-moving nature of a Molniya satellite’s passage through the beam.

Thus, both inclined geostationary satellites and Molniya-type satellites fail to account for the \wow. The former cannot reach the necessary declination to be seen at $-27^\circ$, and the latter cannot mimic the slow, steady motion through the telescope’s beam exhibited by astronomical sources. These orbital dynamics strongly argue against any known satellite type being responsible for an active or passive (reflection) signal detected on August 15, 1977.

A continuous signal from an orbiting satellite will need to be at an orbit beyond the Moon to reproduce the pattern. Although many satellites in operation during that period transmitted within the L-band, albeit not precisely at the frequency of the hydrogen line, their proximity to Earth would have resulted in the emission of signals that were significantly shorter in duration, if they were detectable at all. In contrast, space and planetary missions, positioned far enough to appear stationary, operated within the S or X bands.

\subsection{Solar Activity}

We explored the possibility that the \wow was somehow connected with extraordinary solar activity. Conveniently, the GOES satellites commenced operations in the mid-1970s, and the study of space weather was already established, resulting in a substantial amount of available data. The key conclusion drawn is that solar and geomagnetic activity was notably low during this period.

In mid-August 1977, the Sun was in the rising phase of Solar Cycle 21. The solar minimum had occurred in 1976, so the overall activity was very low. A modest number of sunspot groups were visible on the solar disk during 12–15 August 1977. The international sunspot number was on the order of a few tens. For example, on 14 August 1977 the sunspot number was around 40–42, and on 15 August it was in the upper 30s\footnote{\url{https://www.ngdc.noaa.gov/stp/space-weather/solar-data/solar-indices/sunspot-numbers/depricated/international/tables/daily-sunspot-numbers/daily-sunspot-numbers_1977.txt}}. These values, although higher than the near-zero values seen at the cycle minimum, are very low.

Solar flares may be associated with radio emission. The available records (from NOAA's Solar Geophysical Data and the H$\alpha$ flare event listings) show that no major solar flares (such as M-class or X-class X-ray flares) occurred between 12 and 15 August 1977\footnote{\url{https://www.ngdc.noaa.gov/stp/space-weather/solar-data/solar-features/solar-flares/h-alpha/events/1977/flare-events_1977.txt}}. Only small H-alpha flares, often classified as subflares, were observed during this interval. These subflares are minor brightenings in the chromosphere, usually associated with small sunspots.

Solar radio burst events are powerful radio emissions from the Sun. They are not all well understood and are not always connected with flares. They occur in six distinct categories, labeled as Types~I to~V, with~IV being in turn subdivided between ``stationary'' and ``moving'' \citep[for a review, see ][]{1998ARA&A..36..131B}. Types IV (both stationary and moving) and V can be ruled out because they are broadband in nature. Types~I and~II manifest with frequencies below 500~MHz, which is much lower than that of the \wow signal. Only Type~III bursts are both narrowband and compatible with the \wow signal frequency. However, their typical bandwidth is still too broad ($\sim 1\%$) and, more importantly, they drift very quickly to low frequencies in a matter of seconds. In summary, none of the solar radio burst types matched the characteristics of the \wow signal.

\subsection{Internal Artifacts}

We also considered that the signal might be attributed to an abrupt variation in the telescope gain. Such occurrences were observed in the continuum telescope data from the Ohio SETI LOBES project during the 1990s. To explore this possibility, we simulated the response of the N50CH software under different signals using Equation \ref{eq:snr} to calculate the SNR. Fluctuations in gain can indeed produce a signal, but the resulting pattern is significantly wider and different in shape (with a flat top) than the beam pattern. Furthermore, such gain changes typically impact all frequencies equally, rather than isolating a single frequency channel as noted in the \wow.

The application of Extreme Value Theory (EVT) facilitates the determination of the maximum expected value $\overline{\text{SNR}}_{max}$ for the SNR of a sample $n$ assuming Gaussian noise. Typically, this is estimated using $\log(n)$ or $\sqrt{2\log(n)}$. However, these assessments are upper limits and tend to overstate the actual $\overline{\text{SNR}}_{max}$. A two-term asymptotic expansion pertinent to the half-normal distribution is can be derived as
\begin{equation}
    \label{eq:max}
    \overline{\text{SNR}}_{max} = \sqrt{2 \ln(2n)} - \frac{\ln(\ln(2n)) + \ln(4\pi) - 2\gamma}{2\sqrt{2 \ln(2n)}}
\end{equation}
where $\gamma$ denotes Euler's constant \citep{DASGUPTA201440}. As an illustration to validate Equation \ref{eq:max}, for $n = 7200$, which corresponds to the number of measurements recorded within a single day at a specific declination, $\overline{\text{SNR}}_{max} = 4$. For a data set spanning five years of daily observations, the value reaches 5.5 sigma. This number aligns with empirical data, as random signals seldom exceed a 6-sigma Signal-to-Noise Ratio (SNR).

The performance metrics of the 50 channels of the N50CH receivers did not demonstrate significant discrepancies for channel 2 compared to the other channels. In particular, channel 6 exhibited suboptimal performance, whereas channel 7 showed increased sensitivity. The likelihood that the \wow signal's beam pattern was a random event within the system is negligible. On that particular day, none of the channels exhibited a signal exceeding a 7-sigma level, and the \wow was unique among all the data.

\section{Results and Discussion\label{sec:discussion}}

The revised properties for the \wow from this study are shown in Table \ref{tab:wow}. The updated location of the \wow is 7.25 arcminutes from previous estimates and more than three times narrower (Figure \ref{fig:wowmap}). This adjustment might improve the likelihood of future efforts to identify its source. This new position may also explain the previous challenges in detecting or identifying potential sources. There were many attempts to detect the \wow, but they were broad enough to encompass these new locations, although not necessary with the same sensitivity \citep{1994Icar..112..485G,gray_vla_2001,gray_search_2002,harp2020ata,perez2022breakthrough}. However, some efforts to identify potential sources were concentrated in different areas \citep{2022IJAsB..21..129C,Gontcharov2011}. In this investigation, we have not yet refined the pointing declination of the Big Ear telescope, but that would not reduce the substantial $\pm$20 arcminutes uncertainty linked with the large beam size in declination.


\begin{deluxetable*}{lcc}
\tablecaption{Revised Properties of the \wow. \label{tab:wow}}
\tablehead{
\colhead{Parameter} & \colhead{Previous Values\tablenotemark{a}} & \colhead{Arecibo Wow!}}
\startdata
\noalign{\vskip 1ex}
Date & \makecell[c]{1977 Aug 15 22:16:01 EST \\ 1977 Aug 16 03:16:01 UTC} & \makecell[c]{1977 Aug 15 22:16:06 EST \\ 1977 Aug 16 03:16:06 UTC} \\
Frequency & 1420.455 $\pm$ 0.005\,MHz & 1420.726 $\pm$ 0.005\,MHz \\
Duration & $\geq$ 72 s & $\geq$ 73.4 $\pm$ 0.5 s\\
Signal Strength (SNR) & 30.5 $\pm$ 0.5 & 30.1 $\pm$ 0.4 \\
Flux Density & 54 or 212 Jy & $\geq$ 256 $\pm$ 63\,Jy \\
Radial Velocity (VHEL) & --- & -84 $\pm$ 1 km/s \\
LSR Velocity (VLSR) & --- & -74 $\pm$ 2 km/s \\
Source Size & --- & $\leq$ 1.9 $\pm$ 0.1 arcmin\\
Coordinates (J2000) \\
\quad Positive (East) Horn & \makecell{19:25:31~$\pm$~10\,s \\ -26:57~$\pm$~20\,min} & \makecell{19:25:02~$\pm$~3\,s \\ -26:57:18~$\pm$~20\,min} \\[2ex]
\quad Negative (West) Horn & \makecell{19:28:22~$\pm$~10\,s \\ -26:57~$\pm$~20\,min} & \makecell{19:27:55~$\pm$~3\,s \\ -26:57:13~$\pm$~20\,min} \\
Coordinates (Galactic) \\
\quad Positive (East) Horn & \makecell{$11.65 \pm 0.02\,^\circ$  \\ $-18.89 \pm 0.04\,^\circ$} & \makecell{$11.62 \pm 0.02\,^\circ$ \\ $-17.85 \pm 0.04\,^\circ$} \\[2ex]
\quad Negative (West) Horn & \makecell{$11.90 \pm 0.02\,^\circ$  \\ $-19.48 \pm 0.04\,^\circ$} & \makecell{$11.87 \pm 0.02\,^\circ$ \\ $-19.42 \pm 0.04\,^\circ$} \\[2ex]
\enddata
\tablenotetext{a}{These values were calculated between 1997 and 1998 by \cite{ehman1998big}.}
\end{deluxetable*}

The \wow was more intense than previously believed. Previous estimates of its flux density were 54 Jy or 212 Jy, with 54 Jy considered the most plausible value. We estimated that the flux density was over 250 Jy. Given that the source was not necessarily perfectly centered in the telescope's beam, its actual intensity was likely greater. Few astrophysical radio sources emit at such high-intensity levels. Consequently, this event was exceptional, and it is possible that a fraction of its intensity (from sidelobes) was detected by other radio telescopes worldwide if any were operating at similar frequencies in the Western Hemisphere during that period. That might include the Eastern Hemisphere if the signal lasted longer than 24 hours before or after its detection.

Our recalibration adjusts the frequency of the Wow! Signal to 1420.726 MHz, indicating a significantly higher radial velocity of -84 km/s toward the Sun (blueshift), or -74 km/s in the Local Standard of Rest (LSR) reference frame. This velocity is generally inconsistent with the motion of a source with the galactic rotation. However, there exist known stellar bodies with comparable velocities within the proposed regions, as evidenced by at least three that we identified in the \textit{Gaia} data. Such velocity is also consistent with the relative motion of planetary bodies or gas either moving around or being accreted by stellar objects. Given the multiplicity of scenarios, a thorough process is required to systematically rule out each potential source. In any case, an accurate frequency for the \wow constrains the identification of potential sources and aids in developing search strategies.

The velocity associated with the \wow is also compatible with intermediate-velocity \HI clouds (IVCs). These are concentrations of \HI gas in the Galactic halo or thick disk with LSR velocities typically in the range of about 25 to 90 km/s, \textit{i.e.}, higher than the velocities of most Galactic disk \HI ($\approx$0 km/s), but lower than those of high-velocity clouds (HVCs) ($>$90 km/s) \citep{2022MNRAS.513.3228L}. HVCs, in contrast, show much larger departures from rotation, sometimes exceeding $\pm$300 km/s, and are often associated with accretion from the intergalactic medium or tidal streams. Thus, IVCs occupy a distinct kinematic regime between the disk and HVC populations, representing a transitional component of the Milky Way’s gaseous environment with potential origins in both Galactic and extragalactic processes.

We confirmed that compact \HI clouds generated some signals analogous to the extraterrestrial communication signals that the Ohio SETI project was looking for (\textit{e.g.}, the Wow2 and Wow3 signals described in Section \ref{sec:frequency}). This types of signals were first noticed by the Ohio SETI team, even before the detection of the \wow \citep{1977Icar...30..267D}. They are defined by a narrowband bandwidth of 10 kHz that coincides with the beam pattern, thereby resembling point sources. The signals demonstrate repetition, a trait anticipated for \HI clouds, although they may also exhibit intermittency dependent on variable noise levels. Repetition and intermittency are also expected properties of potential extraterrestrial communications. We conducted comparisons with \HI column density maps and confirmed their identity as \HI clouds. Current SETI initiatives do not have this ambiguity because they operate at a better spectral resolution, sensitivity, and spatial resolution than was possible in the early days of SETI.

There exists no substantiated evidence to suggest that the \wow was caused by interference from external radio frequencies. Terrestrial satellites do not match its duration as they traverse the sky too rapidly. At the time of observation, there were no active planetary missions operating within the L-band frequency range, nor were any significant celestial bodies, including the Moon, situated near the observation beam that could have reflected terrestrial emissions. The frequency associated with \wow does not align with the velocity characteristics anticipated from any object that emits precisely at the hydrogen line frequency within the Solar System. However, this does not rule out transmissions that occur at similar 1420.726 MHz frequencies.

\section{Conclusion\label{sec:conclusion}}

In this study, we made significant revisions to the properties of the Wow! Signal, including its location, intensity, and frequency (Table \ref{tab:wow}). We validated these results by cross-referencing the Ohio SETI data with astronomical databases. Together, these revised properties might help to pinpoint the source or provide new explanations for its origin. We found no evidence that the \wow was induced by external or internal interference of radio frequency.

Our observations align with an astrophysical origin for \wow, as proposed in the recent literature by \cite{mendez2024arecibo}. An emission from a maser flare or superradiance burst from small, cold \HI clouds, offers a plausible explanation for the observed signal. The new properties introduced in this study may serve to either corroborate or refute this hypothesis. In any case, the \wow remains a phenomenon that requires further investigation.

This work presented significant challenges, particularly in the interpretation of decades-old data lacking comprehensive documentation. The expertise and assistance of scientists and long-time volunteers from the Ohio SETI project over the past year were instrumental in clarifying the operational details of the Big Ear telescope. Nevertheless, this study represents only a partial examination of the available material; substantial portions of the Ohio SETI dataset, spanning multiple decades of observations, remain to be analyzed.

An especially noteworthy aspect of this study is its historical context. The Big Ear radio telescope was operated by volunteers for more than 25 years, many of whom remain in contact today through a mailing list. In exploring potential historical connections, we examined the cultural and scientific environment before and after the detection of the \wow in August 1977. Its first known mentions appeared in local newspapers in Ohio. First, in a brief note in the \textit{Dayton Daily News} on January 5, 1979 where the signal was mentioned but not named. Then, the \textit{Lima News} published in 19 July 1979 a feature article on the Big Ear, which mentioned the \wow by name for the first time, although without generating significant public attention. It was in summer 1979 that it also appeared in an article of the third volume of the \textit{Cosmic Search} journal, including a picture of the famous page.

\textit{Cosmic Search} was a unique journal from 1979 to 1982 in which many prominent figures, such as Frank Drake and Carl Sagan, documented and debated the early efforts of SETI. The article on the \wow inspired many to learn more about the signal, including Robert H. Gray, an amateur radio astronomer who went on to become one of its foremost investigators. Over several decades, Gray conducted numerous observations in search of a repetition of the \wow, employing both personal equipment and professional radio telescopes. His sustained efforts and analyses made him a well-known figure within the SETI community, recognized for his substantial contributions to the field. Gray’s dedication to the study of \wow continued until his passing in December 2021. He was among the volunteers committed to understand this remarkable event.

We started the Arecibo Wow! project to search for signals similar to the \wow in archived Arecibo Observatory data. The \wow served primarily as an inspiration for our SETI work; we had no intention of proposing an explanation for the original event, nor of reevaluating its properties, as we believed that all data from the Big Ear telescope had been lost long ago. We are therefore deeply grateful to the scientists and volunteers of the Big Ear (RadObs, as they often call it) for preserving this material for decades.

The Arecibo Wow! team intends to systematically investigate the Ohio SETI data, emphasizing the identification of both narrowband and broadband transient radio signals. Within this dataset, there exist signals whose origins remain undetermined. The objective is to develop an exhaustive archive and catalog of the data, which will be important for studies in time-domain astronomy. In addition, there is a commitment to preserve the legacy of the Big Ear along with that of the Arecibo Observatory. This ambitious effort will entail considerable effort, which will require the commitment once again of numerous volunteers who are driven by a collective enthusiasm for astronomy.

\begin{acknowledgments}
We thank Robert Dixon (formerly Big Ear Deputy Director), Marc W. Abel (Wright State University), Russ Childers (Ohio State University), and Scott Horn (Shorn Media) for their support and providing the Big Ear data. This work was supported by the Center for Advanced Radio Sciences and Engineering (CARSE) (NSF AST-2132229), the NASA Science Activation Program, and the Planetary Habitability Laboratory (PHL) of the University of Puerto Rico at Arecibo.
\end{acknowledgments}


\facility{Big Ear}

\software{\href{https://www.naic.edu/arecibo/}{The Arecibo Observatory Software Library}, \href{https://github.com/wlandsman/IDLAstro}{IDLAstro}, \href{http://cow.physics.wisc.edu/~craigm/idl/idl.html}{Markwardt IDL Library} \citep{2009ASPC..411..251M} \href{https://www.astropy.org/}{Astropy} \citep{astropy:2013, astropy:2018, astropy:2022}, \href{https://aladin.cds.unistra.fr/}{Aladin Sky Atlas}, \href{https://github.com/tesseract-ocr/tesseract}{Tesseract} \citep{10.5555/1288165.1288167}.}

\bibliography{awowii-v1}{}
\bibliographystyle{aasjournal}

\end{document}